\documentclass{aa}  

\usepackage{graphicx}
\usepackage{txfonts}
\usepackage{lipsum}
\usepackage{subcaption}         
\usepackage{lscape}             
\usepackage{placeins}           
\usepackage{color}
\usepackage{siunitx}
\usepackage[version=4]{mhchem}
\usepackage{amsmath}
\DeclareSIUnit\angstrom{\text {Å}}
\usepackage{natbib}
\bibpunct{(}{)}{;}{a}{}{,}

\begin{document}

   \title{SN\,2023fyq: Direct detection of a Type~Ibn supernova progenitor and its multiwavelength environmental constraints}

   \author{
   Xinyi Hong\inst{1, 2}\email{hongxinyi21@mails.ucas.ac.cn}
   \and Ning-Chen Sun\inst{1, 2, 3}\email{sunnc@ucas.ac.cn}\corrauth{sunnc@ucas.ac.cn}
   \and Yali Shao\inst{4}\email{yshao@buaa.edu.cn}
   \and Ke Wang \inst{5, 6}\email{kwang.astro@pku.edu.cn}
   \and Junjie Wu \inst{2, 1}\email{wujj@bao.ac.cn}
   \and Qiang Xi \inst{1, 2}\email{xiqiang23@mails.ucas.ac.cn}
   \and Yi-Han Zhao \inst{1, 2}\email{zhaoyihan20@mails.ucas.ac.cn}
   \and Justyn R. Maund \inst{7}\email{Justyn.Maund@rhul.ac.uk}
   \and Xiaohan Chen \inst{2, 8}\email{xhchen@bao.ac.cn}
   \and Anyu Wang \inst{1, 2}\email{wanganyu22@mails.ucas.edu.cn}
   \and Linxi Zhang \inst{1, 2}\email{zhanglinxi25@mails.ucas.ac.cn}
   \and Jifeng Liu \inst{2, 1, 3, 9}\email{jfliu@nao.cas.cn}
   }

   \institute{
   School of Astronomy and Space Science, University of Chinese Academy of Sciences, Beijing 100049, China
   \and National Astronomical Observatories, Chinese Academy of Sciences, Beijing 100101, China
   \and Institute for Frontiers in Astronomy and Astrophysics, Beijing Normal University, Beijing 102206, China
   \and School of Space and Earth Sciences, Beihang University, Beijing 100191, China
   \and Kavli Institute for Astronomy and Astrophysics, Peking University, Beijing 100871, China
   \and Department of Astronomy, School of Physics, Peking Unversity, Beijing 100871, China
   \and Department of Physics, Royal Holloway, University of London, Egham, TW20 0EX, United Kingdom
   \and School of Physics and Astronomy, China West Normal University, Nanchong 637002, China
   \and New Cornerstone Science Laboratory, National Astronomical Observatories, Chinese Academy of Sciences, Beijing 100012, People's Republic of China
   }

  \abstract 
   {Type~Ibn supernovae (SNe) are characterized by narrow helium emission lines that arise from ejecta-circumstellar medium interaction, yet their progenitors remain debated, with both massive Wolf-Rayet stars and low-mass helium stars in binaries proposed.}
   {We aim to directly identify the progenitor of the Type~Ibn SN\,2023fyq and to characterize its environment in order to constrain the progenitor's nature and evolutionary channel.} 
   {We search for the SN progenitor based on pre-explosion and late-time HST and JWST images and derive its properties by fitting the spectral energy distribution. We investigate the SN environment by probing the stars, dust, ionized gas and molecular gas with a multiwavelength dataset including HST and JWST imaging, VLT/MUSE integral-field-unit spectroscopy, and ALMA CO (2--1) radio interferometry.}
   {We discover a pre-explosion source at the SN position, which is consistent with a hot ($T>15000$\,K) and luminous (log($L$/$L_\odot$)~$\gtrsim$~5.5) SN progenitor and a possible host star cluster. The progenitor is confirmed to have disappeared after the explosion. Analysis of the SN environment implies that the progenitor likely has an age of log($t$/yr) = 7.1--7.2. These phenomena disfavor a very massive single-star progenitor and instead support a binary scenario involving a low-mass helium star and a compact object; the observed progenitor emission likely arises from binary interaction that began at least $\sim$12~yr before the explosion.}
   {SN\,2023fyq is the first Type~Ibn SN with a directly detected progenitor and a possible host star cluster. It adds to the diversity of Type~Ibn SNe in terms of their progenitor channels and mass-loss mechanisms.}

   \keywords{supernova --
                supernova progenitor --
                environmental analysis
               }

   \titlerunning{SN\,2023fyq: Direct detection of a Type~Ibn supernova progenitor and its multiwavelength environmental constraints}
   \authorrunning{Hong et al.}
   \maketitle
\nolinenumbers

\section{Introduction}

Supernovae (SNe) are violent explosive events that denote the terminal phase of the life of stars. They serve as a crucial driving force for cosmic nucleosynthesis, galactic chemical evolution, and energy feedback to the interstellar medium. The SNe can be classified into several distinct types according to their spectral characteristics.
The Type~Ibn classification was first proposed by \cite{2007Pastorello}, with SN\,1999cp recognized as the first confirmed Type~Ibn SN \citep{2000Matheson}. 
The defining feature of this type is the presence of narrow helium emission lines in their spectra with no to weak hydrogen lines. These narrow emission features are considered to arise from the interaction between SN ejecta and circumstellar material (CSM) expelled by the progenitor star shortly before the explosion \citep{2017Hosseinzadeh, 2008Pastorello}. Therefore, the study of Type~Ibn SNe is of vital importance in revealing the late mass-loss processes of hydrogen-poor stripped stars.

Due to the efficient conversion of extensive kinetic energy into radiative energy through CSM-interaction, Type~Ibn SNe are typically more luminous than normal stripped-envelope SNe (Types IIb, Ib, Ic). Their absolute peak magnitudes generally span from $-$17 to $-$20 mag \citep{2017Karamehmetoglu, 2024Wang, 2025Wang}. Meanwhile, the light curves of SNe Ibn are relatively fast-evolving, with rise times of $\sim$10 days and decline rates of $\sim$0.1 mag/day \citep{2025Farias, 2026Cai}. 
The light curves of Type~Ibn SNe show a surprisingly high degree of homogeneity in contrast to the hydrogen-rich CSM-interacting Type~IIn SNe, whose light curves are much more heterogeneous \citep{2017Hosseinzadeh, 2017Smith}. 

The traditional view holds that Type~Ibn SNe originate from massive ($>$25~$M_{\odot}$) Wolf-Rayet (WR) stars or stars that transition from the luminous blue variable (LBV) phase to the WR phase \citep{2007Foley}.
However, WR stars in the Milky Way have never been observed to produce giant outbursts, so \cite{2007Pastorello} and \cite{2007Foley} proposed that the outburst seen two years before SN\,2006jc could arise from a binary companion star of the progenitor.
Recent studies have shown that Type~Ibn SNe are not always found in star-forming regions \citep{2019Hosseinzadeh} or have a relatively low ejecta mass \citep{2025Farias}, implying the existence of an alternative evolutionary channel involving low-mass binary systems. For SN\,2015G, the progenitor was not detected on pre-explosion images, and the derived upper limit ruled out a single massive WR star \citep{2017Shivvers}. For SN\,2006jc, \cite{2016Maund} and \cite{2020Sun} reported a binary companion detection of its progenitor, strongly suggesting a not so massive progenitor in a binary system.
Also, \cite{2022Dessart} conducted detailed spectral synthesis of several Type~Ibn SNe and proposed that the progenitor could be a low-mass helium star in a binary system. Collectively, this growing body of observational and theoretical evidence indicates that the progenitors of Type~Ibn SNe may indeed be low-mass binaries instead of single massive WR stars.
However, it remains an unresolved question whether there is any diversity in the progenitor channels of Type~Ibn SNe and what mechanisms drive their violent pre-explosion mass loss. Until now, apart from the outburst of SN\,2006jc, no progenitors have ever been directly detected in pre-explosion images for any Type~Ibn SNe.

Discovered on April 17, 2023 (JD = 2460051.81897), SN\,2023fyq is a Type~Ibn SN in the galaxy NGC\,4388 \citep{2023De} with a distance of 18~$\pm$~3.7~Mpc (distance module $\mu$ = 31.28~$\pm$~0.45~mag, \citealp{2016Tully}). SN\,2023fyq was initially classified as a Type Ib-pec (peculiar Type Ib) SN \citep{2023Valerin}, but later was identified as a Type~Ibn SN based on its luminosity and spectral evolution \citep{2024Brennan}. 
 Several works have conducted extensive multiband photometric and spectroscopic follow-up observations. 
As noted by \cite{2024Dong}, the evolution of SN\,2023fyq can be divided into four critical phases. From around $t=-$1300 to $-$100\,d relative to the $r$-band maximum, SN\,2023fyq exhibits relatively steady precursors of $\sim$20 mag. From $t=-$100\,d, the precursor emission gradually brightens to $\sim$16 mag until SN\,2023fyq exploded at $t=-$11\,d. The "main" SN light curve shows a double-peaked profile, with the first peak powered by a shock breakout (SBO) and the second by ejecta-CSM interaction. At the second peak, SN\,2023fyq reached its $r$-band maximum of $\sim$13 mag. From $t=$20 days after the maximum, the SN light curve shows a linear tail powered mainly by radioactive decay. 

In addition, \citet{2024Dong} found that there were still strong medium-width helium lines in the nebular spectrum of SN\,2023fyq at $\sim$136.8\,ds, indicating that the ejecta-CSM interaction remained active in the late phase after the explosion.
\cite{2025Baer-Way} reported the first radio view of a Type~Ibn SN in SN\,2023fyq. They fit the radio emission from 58-185 days with a CSM of density $\sim10^{-18}$~g/cm$^3$ at a radius of $10^{16}$~cm, corresponding to a mass-loss rate of $\sim4~\times~10^{-3}~M_\odot$/yr at 0.7-3 years before the explosion. Meanwhile, their late-time (at 525 days) non-detections in both radio and X-ray bands suggested a dense shell-like CSM with a width of at most 4~$\times~10^{15}$ to 2~$\times~10^{16}$~cm.

These observations suggest that the progenitor of SN\,2023fyq was extremely unstable before the explosion with a violent mass-loss process. 
\cite{2024Dong} proposed that the progenitor was likely a binary system comprising a low-mass helium star and a compact object. The precursor emission was attributed to mass transfer driven by the expansion of the helium star; during this phase, an equatorial accretion disk may have formed with a mass of approximately 0.6 M$_\odot$. The ultimate explosion was possibly triggered either by core collapse of the helium star or by merger-induced instability of the binary system.
\cite{2024Tsuna} also established a binary merger model of a low-mass helium star and a neutron star companion with an initial orbital period of 100 days, which successfully matched the observations of SN\,2023fyq. The radio data from \cite{2025Baer-Way} are also consistent with the predictions of this merger model.

In this work, we report the direct detection of the progenitor of SN\,2023fyq, as well as a possible host star cluster, based on images taken by the \textit{Hubble} Space Telescope (HST) and the \textit{James Webb} Space Telescope (JWST). We note that K. Taggart et al. have independently discovered a progenitor candidate for SN\,2023fyq and reported it at the \textit{Transients From Space} workshop\footnote{\url{https://www.stsci.edu/contents/events/stsci/2025/march/transients-from-space}; \url{https://www.youtube.com/watch?v=NpIV632xgR8}.}. However, at that point, it was not clear whether the candidate had disappeared after the explosion; furthermore, no published study has systematically analyzed its progenitor and environment. As will be shown in this paper, we find that at late times SN\,2023fyq has become much fainter than the progenitor candidate; this confirms that the candidate is the genuine progenitor and has disappeared after explosion. A study of the SN environment can also provide key insights into their progenitors (e.g., \citealp{2023Sun2, 2025Xi, 2026Xi, 2026Niu} ). Therefore, we perform a detailed analysis of the host galaxy and local environment of SN\,2023fyq based on HST and JWST imaging, integral field unit (IFU) spectroscopy by the Multi Unit Spectroscopic Explorer (MUSE) on the Very Large Telescope (VLT), and CO (2--1) radio interferometry by the  Atacama Large Millimeter/sub-millimeter Array (ALMA). Our aim is to reveal and understand what progenitor gives rise to this SN and to derive its key parameters.

As mentioned, all epochs in this paper are relative to the $r$-band maximum of SN\,2023fyq at JD = 2,460,154.3 $\pm$ 0.5 \citep{2024Dong} unless otherwise specified.
We use a Galactic reddening of $E(B-V)_{\rm MW}$ = 0.029~mag and a host-galaxy distance of 18~$\pm$~3.7~Mpc, both consistent with \cite{2024Dong}.
This paper is organized as follows. Observations we used are described in Section~\ref{sec:data}. We search for and analyze the progenitor in Section~\ref{sec:progenitor}, and investigate the SN environment in Section~\ref{sec:env}. Finally, we summarize and discuss the main results in Section~\ref{sec:sum}.

\section{Data}\label{sec:data}

\subsection{HST and JWST imaging}

In this work, we used HST and JWST images, the complete information for which is presented in Table~\ref{tab:data}. The HST observations were conducted by the Wide Field Camera 3 (WFC3) in both the Ultraviolet-Visible (UVIS) channel and the Infrared (IR) channel, spanning a wide wavelength from the F336W band to the F160W band. Most of the observations were conducted in 2011, that is,  $\sim$12\,years before the explosion of SN\,2023fyq, except one F336W observation that was performed late in 2025, that is,  $\sim$2\,years post-explosion.

The SN site was also observed by JWST with the Near-Infrared Camera (NIRCam) and the Mid-Infrared Instrument (MIRI). 
The MIRI observations were conducted in the F560W, F1000W, F1500W, F1800W, F2100W, and F2550W filters at two epochs in June 2023 and June 2024, corresponding to $t=-$31 days (during the outburst phase before core collapse) and $t$ = 321\,days relative to the $r$-band maximum of SN\,2023fyq; we use the F1800W images to locate the SN position (see Section~\ref{sn_position.sec}) while those in the other filters are not used in this work. The NIRCam observations were taken in the F182M and F335M filters in January 2025, or $t$ = 531\,ds after the $r$-band maximum. We use the NIRCam images to study the SN environment.

All the above data were obtained from the Mikulski Archive for Space Telescopes (MAST)\footnote{\url{https://mast.stsci.edu/portal/Mashup/Clients/Mast/Portal.html}}. 
Photometry was performed with \textsc{dolphot} \citep{2000Dolphin, 2016Dolphin}. 
For source detection, we used the drizzled image of each filter as a reference and analyzed each filter separately. The S/N thresholds adopted later in this work are, therefore, defined on a per‑filter basis.
We used \texttt{FitSky = 2} and \texttt{RAper = 3}, which are recommended by the user manual for crowded regions. We set \texttt{Force1 = 1} (force all objects to be fitted as stars) and \texttt{ApCor = 1} (turn-on aperture correction) for better point-source photometry. 
For HST/WFC3 we used \texttt{UseWCS = 1} to perform image alignment based on the header information of the world coordinate system (WCS) while for JWST/NIRCam  we used \texttt{UseWCS = 2} since the \textsc{dolphot/nircam} module does not include the default distortion parameters.
In addition, we used \texttt{WFC3useCTE = 0} for HST/WFC3/UVIS and \texttt{WFC3useCTE = 1} for HST/WFC3/IR because the images of the former have already been corrected for charge transfer inefficiency at the pixel level, while those of the latter have not. 
All other parameters are the same as recommended by the \textsc{dolphot} user manual.

\subsection{VLT/MUSE IFU spectroscopy}

The host galaxy of SN\,2023fyq was observed with VLT/MUSE in March 2022 (Program ID: 108.229J.001; PI: Venturi, Giacomo). The observations were carried out in the wide-field mode (WFM) with an extended wavelength setting, covering a field of view of \SI{1}{\arcmin}\,$\times$\,\SI{1}{\arcmin} at a spatial sampling of 0.2" and a wavelength range \SIrange{4700}{9350}{\angstrom} at a spectral sampling of \SI{1.25}{\angstrom}. The total exposure time was 4440 s with two equal-duration exposures. The raw data were well reduced by the standard pipeline, and we retrieved the calibrated datacube from the European Southern Observatory (ESO) data archive\footnote{\url{https://archive.eso.org/scienceportal/home}}.

We used the \textsc{ifuanal} package\footnote{\url{https://ifuanal.readthedocs.io/en/latest/index.html}} to further process the datacube. 
The datacube was first corrected for Galactic extinction and de-redshifted.
Then, to mask and remove the stellar continuum in the galaxy, the spaxels were binned using Voronoi tessellation to achieve S/N of at least 120 over the wavelength range \SIrange{5590}{5680}{\angstrom}. In each bin, we fitted the spectrum with the \textsc{starlight} package \citep{2005Cid} based on the \cite{2003Bruzual} stellar population models. Therefore, the stellar continuum was able to be masked and subtracted from the spectrum, leaving only the gaseous component in the datacube.

\subsection{ALMA CO (2--1) radio interferometry}

We adopted the ALMA CO (2--1) observations using the 7-m array toward NGC\,4388. These observations were conducted on December 22, 2019 with ten antennas (ID: 2019.1.00763.L) and April 27 and 28, 2024 with 11 antennas (ID: 2023.1.00026.S), respectively. The on-source integration time was 15 and 21 min, respectively. The flux density scale was established using scans of the standard calibrators J1058+0133 and J1256-0547 (ID: 2019.1.00763.L) and 1229+0203 (ID: 2023.1.00026.S). The phase and the water vapor were further checked by observing the nearby calibrators of J1215+1654  (ID: 2019.1.00763.L), and J1222+0413 and J1254+1141 (ID: 2023.1.00026.S). 

We retrieved data from the ALMA science archive\footnote{\url{https://almascience.nrao.edu/}} and performed calibration with the standard pipeline and the \textsc{casa} package\footnote{\url{https://casa.nrao.edu/}}.
We only considered the frequency range that overlapped with our science goals for the final combination of multiexecution data. We made the line datacube from the combined calibrated data using the \textsc{tclean} task with \texttt{Briggs} weighting and \texttt{robust = 0.5} (in order to optimize the sensitivity per frequency bin and the resolution of the final map), and with auto-multithresh mask and \texttt{side-lobethreshold = 2.0}, \texttt{noisethreshold = 4.25}, \texttt{minbeam-frac = 0.3}, \texttt{lownoisethreshold = 1.5}, \texttt{negativethreshold= 15.0}, and \texttt{threshold = 3$\sigma$} in a gridder mode of mosaic. 
The synthesized beam size is 7.4"\,$\times$\,4.4", corresponding to 646\,pc\,$\times$\,384\,pc at the host galaxy's distance. The noise level in a 15.625 MHz channel is $\sim$5.5\,mly\,beam$^{-1}$, and the moment-0 map (i.e., the velocity-integrated intensity) has a sensitivity of $\sim$1\,Jy\,beam$^{-1}$\,kms$^{-1}$.

\section{Progenitor detection \label{sec:progenitor}}

\subsection{Identification and photometry}
\label{sn_position.sec}
\begin{figure}[ht!]
   \centering
   \includegraphics[width=\hsize]{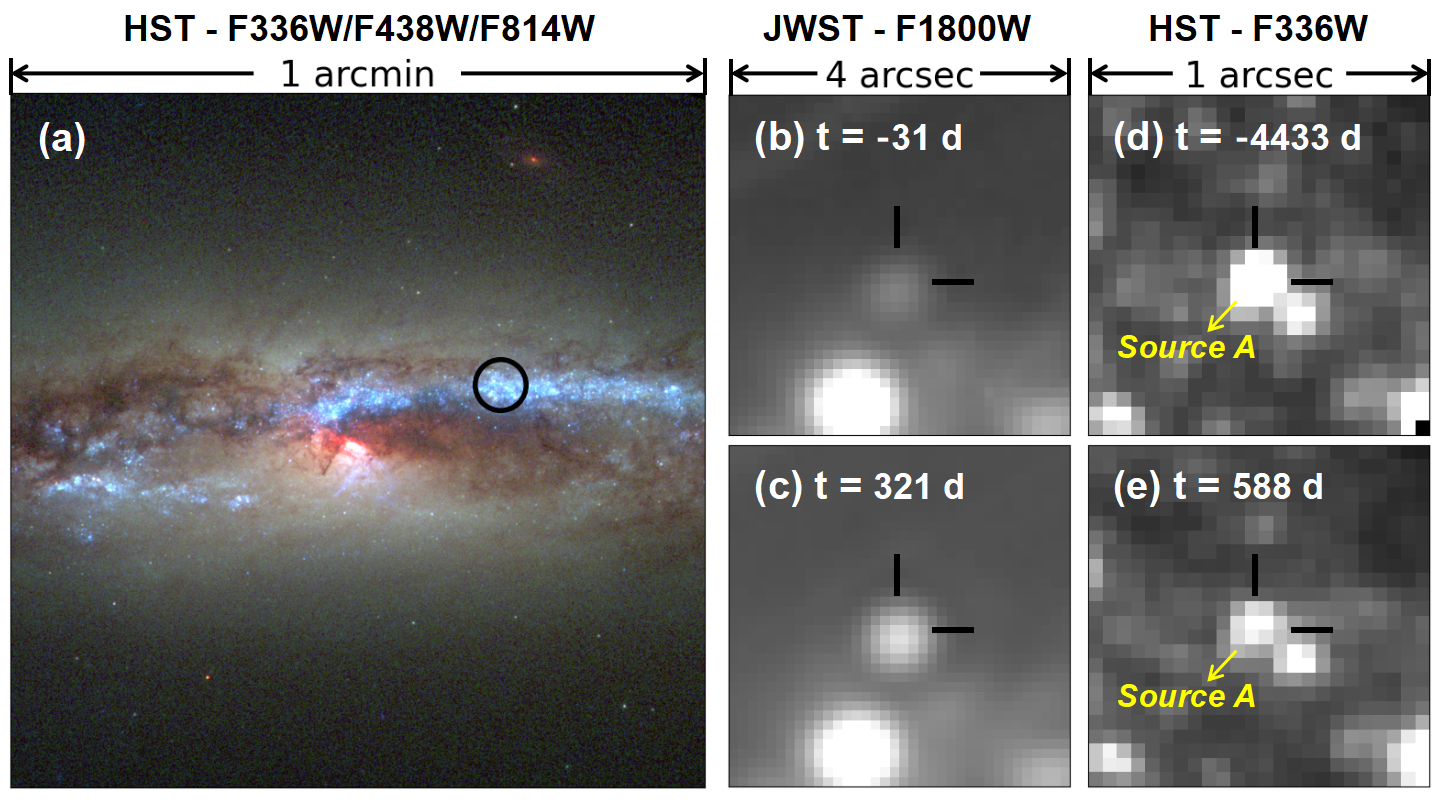}
      \caption{(a) F336W/F438W/F814W three-color composite image of the host galaxy by HST observations. The black circle shows the SN site. (b, c) JWST/MIRI/F1800W images of the SN site during the outburst phase and after the SN explosion. (d, e) Pre-explosion and late-time images taken by HST/WFC3/UVIS in the F336W band. The crosshairs mark the SN position. All images are aligned with north up and east to the left.}
         \label{fig:env}
   \end{figure}

Fig.~\ref{fig:env}(a) shows the HST F336W/F438W/F814W three-color composite image of the host galaxy NGC\,4388. SN\,2023fyq is located on one of the spiral arms at $\sim$10\,arcsec away from the galaxy center. The local SN environment is a very dense field with a large number of bright blue stars and/or clusters.

Fig.~\ref{fig:env}(b,c) compares the JWST/MIRI/F1800W images taken during the outburst phase shortly before core collapse ($t=-31$\,d) and after the explosion of SN\,2023fyq ($t=321$\,d). Near the reported SN coordinates, we find a point source with variable brightness, significantly brighter after the SN explosion. We identify this source as SN\,2023fyq.
We then used the \textsc{photutils} package\footnote{\url{https://github.com/astropy/photutils}} to pinpoint the source positions and computed the coordinate transformation matrices across different images based on common stars; in this way, we determined the precise SN positions on all HST and JWST images. 

In the pre-explosion HST images, a point source (labeled as Source~A) is significantly detected at the SN position in all filters. As an example, Fig.~\ref{fig:env}(d) shows the pre-explosion image in the F336W band. 
At late times, Source~A is still visible on the F336W image (Fig.~\ref{fig:env}e), but appears much fainter than the pre-explosion level, confirming that Source~A is physically associated with SN\,2023fyq.

\begin{figure}[ht!]
   \centering
   \includegraphics[width=\hsize]{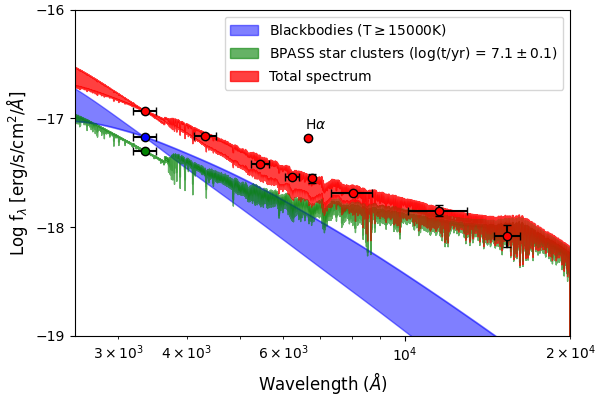}
      \caption{Observed and model pre-explosion SEDs (red) of Source~A and its two components, i.e., a SN progenitor modeled as a blackbody (blue) and a star cluster that persists after the SN explosion (green). The green data point corresponds to the late-time detection of Source~A while the blue data point corresponds to the flux that has vanished from pre-explosion to late-time epochs. The horizontal error bars show the rms widths of the filters. For the F110W and F160W filters, the vertical error bars have accounted for the uncertainty due to contamination of neighboring stars (see Appendix B for details) while for the other filters the vertical error bars reflect photometric uncertainties.}
         \label{fig:sed}
   \end{figure}

Based on photometry with the \textsc{dolphot} package, we obtained the magnitudes of Source~A across multiple wavelength bands (Table~\ref{tab:mag}). We can clearly see that the brightness of Source~A in the F336W band has faded from pre-explosion to late time, with a magnitude difference of ${\Delta}m$ = 0.925 mag. 
Note that the two HST/WFC3/IR images in the F110W and F160W bands have relatively lower spatial resolutions and the detected source is likely contaminated by the close neighbors of Source~A. We used the HST/WFC3/UVIS/F814W and JWST/NIRCAM/F182M images to resolve the stellar environment and estimate the contamination in the F110W and F160W bands with interpolation. The details can be found in Appendix~\ref{app:mixed}. The final spectral energy distribution (SED) of Source~A is presented in Fig.~\ref{fig:sed}.

    \begin{table}[ht!]
    \caption{\label{tab:mag} HST photometry of Source~A.}
    \centering
    \begin{tabular}{ccc}
        \hline\hline
        Filter & Pre-explosion & Late-time \\
        & ($t=-4461\sim-4433$\,d) & ($t=$588\,d) \\
        \hline
        F336W & 21.126 (0.020) & 22.051 (0.036)  \\
        F438W & 22.479 (0.037) & -  \\
        F547M & 22.456 (0.042) & -  \\
        F621M & 22.320 (0.045) & -  \\
        F665N & 21.189 (0.043) & -  \\
        F673N & 22.074 (0.099) & -  \\
        F814W & 21.862 (0.023) & -  \\
        F110W & 20.995 (0.063) & -  \\
        F160W & 20.372 (0.069) & -  \\
        \hline\hline
    \end{tabular}
    \tablefoot{The F110W and F160W photometry is likely contaminated by some close neighbors due to the relatively lower spatial resolution of HST/WFC3/IR. See Appendix B for details. All magnitudes in this paper are in the VEGA photometric system.}
    \end{table}

\subsection{Modeling the SED \label{subsec:sed}}

\begin{figure}[ht!]
   \centering
   \includegraphics[width=\hsize]{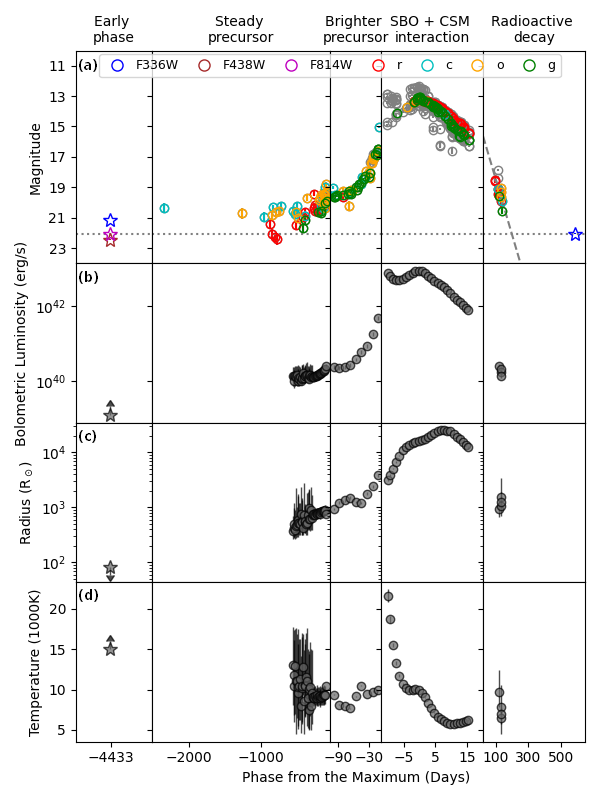}
      \caption{The pre- and post-explosion evolution of the (a) magnitude, (b) bolometric luminosity, (c) blackbody radius and (d) temperature of SN\,2023fyq.
      Stars correspond to detections in this work while circles are data from \cite{2024Dong}. In the magnitude panels, colored circles correspond to different bands as shown in the legend above; gray circles represent magnitudes in other bands from \cite{2024Dong}.
      The full evolution is divided into five critical phases according to \citet{2024Dong} and this work. 
      Our detection is around 2100 d before the first precursor detections by \cite{2024Dong}.}
         \label{fig:dong}
   \end{figure}

Fig.~\ref{fig:dong}(a) compares the pre-explosion and late-time brightness of Source~A with the light curve of SN\,2023fyq as reported by \citet{2024Dong}. Our pre-explosion detection of Source~A corresponds to $t=-4461 \sim -4433$\,d, long before the first precursor detections at $t=-2337$\,d by \citet{2024Dong} using ground-based telescopes. The late-time HST/WCF3/UVIS/F336W detection of Source~A corresponds to $t=$ 588\,d; at this epoch, SN emission should have already faded based on the decline rate of the SN light curve tail. 

Many SNe show late-time emission powered by CSM interaction \citep{2026Fremling}. At a similar phase, \cite{2025Baer-Way} reported no radio detection with a fairly weak upper limit on the mass-loss rate ($\dot{M}~<~2.5~\times~10^{-3}~M_\odot~$yr$^{-1}$), which is only marginally below that at early times ($~\sim4~\times~10^{-3}~M_\odot~$yr$^{-1}$). This limit alone cannot rule out a late‑time SN contribution to the F336W flux. On the other hand, HST UV observations of other Type Ibn SNe show no signs of CSM interaction at late epochs \citep{2020Sun}. Furthermore, as discussed in the next paragraph, there is a young star cluster at the SN position; the late‑time UV brightness is likely from the cluster and under this assumption, the derived cluster age is very consistent with that of the surrounding stellar population (see Section~\ref{subsec:cmd}). 
Therefore, we assume that the SN light had completely faded in the following analysis and discuss later the influence on our results if a CSM interaction exists.

We model the pre-explosion SED of Source~A as composed of an SN progenitor and a star cluster and the late-time SED as the cluster alone, assuming no more late-time radiation from the SN itself and that the progenitor has disappeared after the SN explosion. We suggest the presence of a star cluster for the following reasons.
(1) The F665N flux of Source~A is much higher than the continuum (Fig.~\ref{fig:sed}) and suggests significant H$\alpha$ emission; since the progenitor of SN\,2023fyq is hydrogen-poor, this emission is most likely due to ionized gas associated with a star cluster.
(2) The F110W and F160W bands show a clear IR excess compared with the Rayleigh-Jeans tail extrapolated from bluer bands; this is naturally explained by the integrated emission from red supergiants (RSGs) within a star cluster.
(3) Source~A is located in a very dense stellar field (Fig.~\ref{fig:env}), where physical association or chance alignment with a star cluster could be very common.

We use blackbody models to fit the progenitor and binary population and spectral synthesis (BPASS) models \citep{2022Byrne} to fit the cluster.
Although the SED of stars with intense mass‑loss activity do not perfectly match a blackbody \citep{2018Gotberg}, the blackbody approximation remains a practical model for initial characterization.
We adopt a SN host-galaxy reddening of $E(B-V)_{\rm host}$ = 0.037~$\pm$~0.09~mag as derived by \citet{2024Dong} and an extinction law of $R_V$ = 3.1 \citep{1989Cardelli}. The best-fitting model spectra are displayed in Fig.~\ref{fig:sed}. 
The corresponding parameter ranges were determined manually after extensive testing of a large set of parameter combinations. Although the model includes several free parameters, the IR flux is dominated by the cluster component, which places tight constraints on the cluster age and renders degenerate solutions highly unlikely. We find the cluster to have an age of log($t$/yr) = 7.1~$\pm$~0.1 and an initial mass of log($M$/$M_\odot$)= $3.8^{+0.1}_{-0.1}$; the SN progenitor has a temperature of $>$15000~K, a luminosity of log(L/L$_\odot$)~$\gtrsim$~5.5, and a radius of log(R/R$_\odot$)~$\lesssim$~2.
The high luminosity, combined with the precursor event detected at $\sim$12\,yr before the explosion (Fig.~\ref{fig:dong}a), implies that the progenitor was not in a quiescent state. Consequently, the luminosity derived above should be interpreted as an upper limit to the progenitor's quiescent luminosity.

We also consider the possibility that the star cluster may be in a chance alignment with the SN progenitor and has a different extinction from that derived by \citet{2024Dong} for the SN. As we show later in Section~\ref{sec:env}, there are two stellar populations in the SN environment.
The extinction of the older stellar population is consistent with that of the SN, while the extinction of the younger stellar population is slightly higher. 
We performed another fitting using the extinction of the younger stellar population, and the derived cluster age did not change significantly. 
Furthermore, the cluster age is very consistent with that of the older stellar population; therefore, the star cluster is most likely to be part of the older stellar population with a lower extinction. 
As discussed in Section~\ref{subsec:stru}, we suggest that this star cluster possibly hosts the progenitor of SN\,2023fyq although chance alignment cannot be fully excluded.

\begin{figure}[ht!]
   \centering
   \includegraphics[width=0.85\hsize]{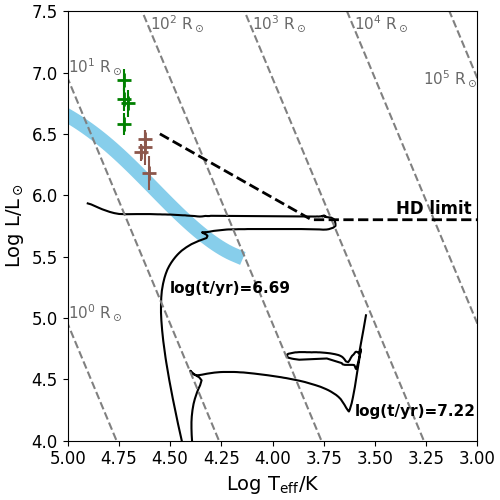}
      \caption{Position of SN\,2023fyq's progenitor on the HR diagram (shaded blue region). The solid black lines are \textsc{parsec} v1.2S single stellar isochrones \citep{2012Bressan}, the reference ages of which are derived by environmental stellar populations fitting (see Section~\ref{subsec:cmd} for details). The dashed line black is the \cite{1979Humphreys} limit for stellar luminosity and the gray dashed lines correspond to constant blackbody radii. The “+" symbols show some of the most luminous stars in the Milky Way (NGC 3603-A1a, A1b, B and C; in brown) and the Large Magellanic Cloud (R136a1–3 and b; in green; \citealp{2010Crowther}).}
         \label{fig:hr}
   \end{figure}

Fig.~\ref{fig:dong}(b--d) compares the blackbody parameters of the SN progenitor with those of SN\,2023fyq from \citet{2024Dong}.
From the early pre-explosion phase to the precursor phase, the luminosity increased from $>$$10^{39}$~erg/s to $\sim$$10^{40}$~erg/s, and the blackbody radius increased from $<$100~R$_\odot$ to $>$400~R$_\odot$, while its temperature decreased from $>$15,000\,K to $<$12,500\,K. Fig.~\ref{fig:hr} displays the possible position of the SN progenitor on the Hertzsprung–Russell (HR) diagram. The progenitor luminosity lies at the upper end of the possible luminosity range for single stars, close to some of the most luminous very massive stars of $>$100~M$_\odot$ in the Milky Way or in the Large Magellanic Cloud \citep{2010Crowther}. However, we deem it unlikely that the progenitor is a very massive star, which will be discussed later.

\section{SN environment \label{sec:env}}

\subsection{Structure of the host galaxy \label{subsec:gal}}

   \begin{figure*}[ht!]
   \centering
   \includegraphics[width=\hsize]{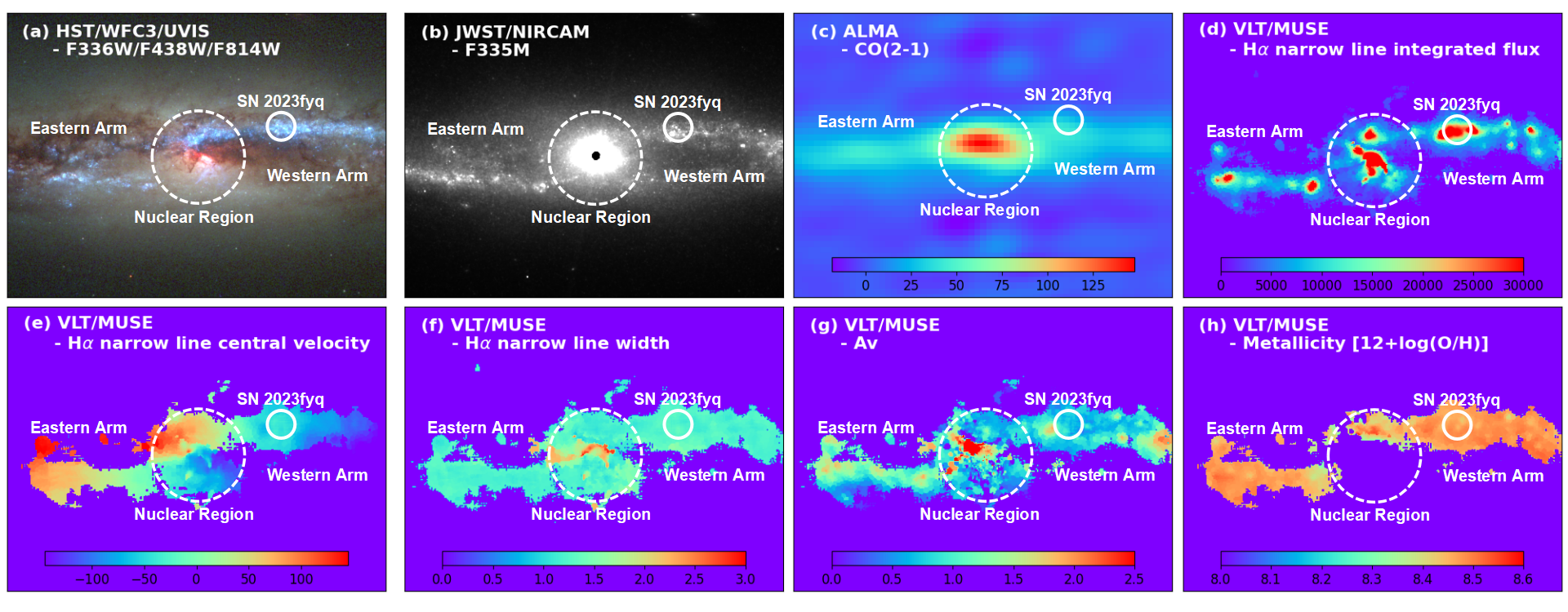}
      \caption{Maps of SN\,2023fyq's host galaxy NGC\,4388. (a) The F336W/F438W/F814W three-color composite image observed by HST/WFC3/UVIS. (b) The F335M-band image observed by JWST/NIRCam. (c) The velocity-integrated intensity map of the CO (2--1) emission line observed by ALMA, in units of $\rm{Jy\,beam^{-1}\,km\,s^{-1}}$. (d) The wavelength-integrated flux map of the H$\alpha$ narrow emission line observed by VLT/MUSE, in units of $\rm{10^{-20}\,erg\,s^{-1}\,cm^{-2}}$. (e) The H$\alpha$ line central velocity map in units of $\rm{km\,s^{-1}}$. (f) The H$\alpha$ line width map in units of \si{\angstrom}. (g) The host extinction map in units of mag. (h) The 12 + log(O/H) metallicity map in units of dex. In (d--h), only parameters of the narrowest component are displayed when there are multiple Gaussian components in the line profile in a spaxel. The solid circles show the location of SN\,2023fyq and the dashed circles mark the nuclear region with the eastern arm and western arm located on either side of it. All panels are aligned with north up and east to the left.}
         \label{fig:galaxy}
   \end{figure*}

With the abundant data from HST, JWST, ALMA and VLT, we can analyze the structure of SN\,2023fyq's host galaxy in detail. It is worth noting that NGC\,4388 has a nearly edge-on inclination of $79^{\circ}$ \citep{2016Damas}. Such a high inclination significantly increases line-of-sight dust extinction and further complicates optical-band analyses.

Fig.~\ref{fig:galaxy}(a) shows the optical image of NGC\,4388 by HST. We can see two spiral arms (labeled as the eastern arm and western arm), composed of blue and bright stars, as well as heavy dust obscuration on the disk. The spiral arms on both sides appear to be highly asymmetric: the western arm is more prominent and continuous, whereas the eastern arm is much fainter and patchy. This asymmetry is likely due to the different relative distributions of stars and dust along the line of sight; on the eastern arm, dust lies in the foreground relative to the stars, whereas the reverse is true for the western arm.
Therefore, the eastern arm is heavily obscured by the foreground dust, while the western arm is much less affected.

Fig.~\ref{fig:galaxy}(b) shows the near-IR F335M image by JWST. At this wavelength, dust extinction is much lower and stars behind dust can be seen more easily. Compared with the optical image, the eastern arm is much brighter and more continuous on the near-IR image, and it is now more symmetric to the western arm. This confirms the speculation in the previous paragraph that foreground dust obscuration causes the faintness and patchiness of the eastern arm on the optical image.

\begin{figure}[ht!]
   \centering
   \includegraphics[width=0.85\hsize]{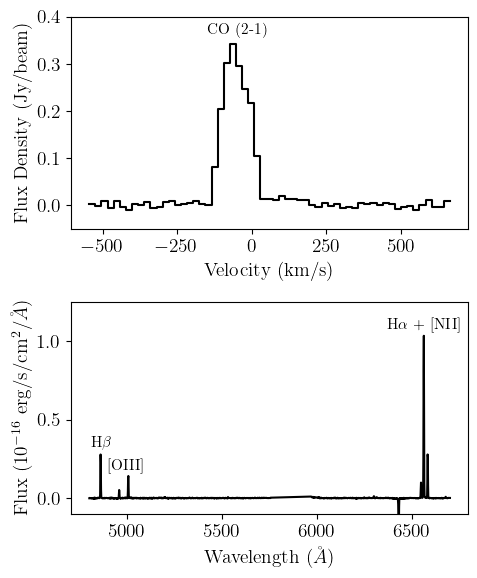}
      \caption{\textit{Top:} Molecular line emission of CO (2--1) at the position of SN\,2023fyq as observed by ALMA; the host galaxy's redshift has been corrected and the blueshift of the line center reflects the galaxy's disk motion toward the observer. \textit{Bottom:} VLT/MUSE spectrum at the SN position with the stellar continuum subtracted; the host galaxy's redshift and Galactic foreground reddening have been corrected.}
         \label{fig:spec}
   \end{figure}

ALMA has detected significant CO (2--1) molecular line emission across the host galaxy. As an example, the top panel of Fig.~\ref{fig:spec} shows the CO (2--1) line in the spectrum at the SN site.
Fig.~\ref{fig:galaxy}(c) presents the velocity-integrated intensity map of the CO (2--1) emission line by ALMA. 
There is a large amount of molecular gas at the galactic center and on the spiral arms. The spatial distribution is very symmetric on both spiral arms, which is similar to the stellar distribution on the near-IR image (Fig.~\ref{fig:galaxy}b) but very different from that on the optical image (Fig.~\ref{fig:galaxy}a).

We further used the VLT/MUSE IFU datacube to analyze the ionized gas. 
As an example, the bottom panel of Fig.~\ref{fig:spec} shows the continuum-subtracted spectrum at the SN position. 
There are prominent emission lines H$\beta$, \ce{[\ion{O}{iii}]}, \ce{[\ion{N}{ii}]}, and H$\alpha$ in the spectrum. 
On the spiral arms, the line profiles can be well described by single-Gaussian profiles. In the nuclear region, however, the line profiles can be much more complicated with more than one, sometimes even up to four, Gaussian components. This indicates that NGC\,4388 has complex gas kinematics around the galaxy center, likely caused by active galactic nucleus (AGN) outflows and/or other nuclear activities \citep{2026Fujiwara}. We fit the lines with single- or multi-Gaussian functions for all 2157 spaxel bins across the galaxy, examples of which can be found in Appendix~\ref{app:specline}.

Fig.~\ref{fig:galaxy}(d-f) shows maps of line integrated flux, central velocity, and width for H$\alpha$; parameters of the narrowest component are adopted in case there are multiple components at a spaxel bin. 
Fig.~\ref{fig:galaxy}(d) shows strong H$\alpha$ emission at the nuclear region and on both spiral arms.
We adopted the Baldwin-Phillips-Terlevich (BPT) diagram to identify whether the emission lines originate from an AGN or star formation (see details in Appendix~\ref{app:bpt}). 
We find that the ionized gas on the spiral arms is excited by star formation, while that in the nuclear region is excited by AGN.

The line central velocity map (Fig.~\ref{fig:galaxy}e) shows a clear pattern of galactic disk rotation. The gas on the eastern arm is moving away from us along the line of sight (redshifted), while that on the western arm is moving toward us (blueshifted). Fig.~\ref{fig:galaxy}(f) shows that most regions on the spiral arms have similar line widths of $\sim$\SI{1.2}{\angstrom}. This width is almost the same as the instrumental broadening \citep{2017Guerou}, indicating a very low intrinsic velocity dispersion of the ionized gas. 

We further used the Balmer decrement to estimate the interstellar extinction (Fig.~\ref{fig:galaxy}g), assuming an intrinsic flux ratio of H$\alpha$/H$\beta$ = 2.86 \citep{2006Osterbrock} and a \cite{2004Fitzpatrick} extinction law with $R_V$ = $A_V/E(B - V)$ = 3.1. 
Overall, we can observe that the extinction on the eastern arm is higher than that on the western arm, which is consistent with our earlier inference about the spatial distribution of objects within the galaxy, and the host extinction at the SN site is $A_V^{\rm host}$ $\sim$ 0.8 mag.

We estimate the gas-phase metallicity via the strong-line diagnostics, using the O3N2 calibration from \cite{2013Marino}. 
Note that this calibration is based on H\textsc{ii} regions instead of AGN-ionized gas. Therefore, we only derive the metallicity for the regions ionized by star formation (Fig.~\ref{fig:galaxy}h).
The spiral arms have a roughly uniform metallicity of 12 + log(O/H) = 8.5 $\pm$ 0.18 dex, which is slightly lower than the solar metallicity (8.69 dex, \citealp{2009Asplund}).
This metallicity value has been utilized in the SED fitting of Source~A (Section~\ref{subsec:sed}) and will be used for the resolved stellar population fitting in the next section (Section~\ref{subsec:cmd}).

\subsection{Stellar populations in the local SN environment\label{subsec:cmd}}

With the high-resolution photometry data from HST, we can perform a quantitative analysis of the star-forming history in the local SN environment. Here we only use the wide filters of F336W, F438W, and F814W. In a circular region with a radius of 100\,pc ($\sim$1 arcsec) around SN\,2023fyq, we selected good stars based on the following criteria:
(1) type of source, \texttt{obj = 1}; 
(2) S/N, \texttt{snr > 5}; 
(3) source sharpness, \texttt{|shp| < 0.5}; 
(4) source crowding, \texttt{crd < 2}; 
(5) photometry quality flag, \texttt{qfg < 3}. 
We finally selected a catalog of 135 sources (excluding the progenitor candidate) for further analysis. 

\begin{figure}[ht!]
   \centering
   \includegraphics[width=\hsize]{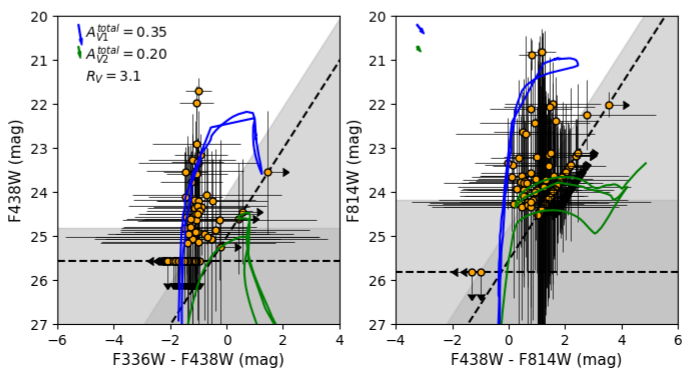}
      \caption{CMDs of all stellar sources in the local environment of SN\,2023fyq. In both panels, the blue and green isochrones correspond to stellar populations with log-ages of log($t$/yr) = $6.69$ and $7.22$, respectively, and have been shifted along the reddening vectors shown by the arrows. The arrows in the upper-left corner, color-matched to each isochrone, represent reddening vectors for $R_V$ = 3.1 and total extinctions of $A_V^{\rm{total}}$ = 0.35 and 0.20 (which correspond to best-fitting host extinctions of $A_V^{\rm{host}}$ = 0.26 and 0.11 for the two populations, respectively, in addition to a Milky Way reddening of $E(B-V)_{\rm{MW}}$ = 0.029 mag).The dashed lines indicate the detection limits, and the gray shaded regions represent the uncertainties in those detection limits for the corresponding filters. Note that the error bars have been multiplied by a factor, estimated by artificial star test, to account for the additional error due to source crowding.}
         \label{fig:cmd}
   \end{figure}

We performed artificial star tests to estimate the detection limits within the SN environment and calculate a scaling factor of the reported photometric uncertainty to account for the additional error introduced by source crowding (the method is the same as that in \citealp{2024Hong}). 
Fig.~\ref{fig:cmd} shows the color-magnitude diagrams (CMDs) drawn with the catalog. It can be seen that the photometric error caused by crowding is very large.

Based on this catalog, we use a hierarchical Bayesian approach (\citealp{2016Maund}; \citealp{2021Sun, 2022Sun, 2023Sun1, 2023Sun2}; \citealp{2023Niu}) to fit the observed environmental stars with model stellar populations (based on \textsc{parsec} v1.2S stellar isochrones\footnote{\url{http://stev.oapd.inaf.it/cgi-bin/cmd}}, \citealp{2012Bressan}). 
We find that the environmental stars can be well fitted with two model stellar populations, one with an age of log($t$/yr) = $6.69^{+0.01}_{-0.01}$ and a host extinction of $A_V^{\rm{host}}$ = $0.26^{+0.09}_{-0.07}$ and the other with log($t$/yr) = $7.22^{+0.02}_{-0.02}$ and $A_V^{\rm{host}}$ = $0.11^{+0.02}_{-0.02}$. 

   \begin{figure}[ht!]
   \centering
   \includegraphics[width=0.7\hsize]{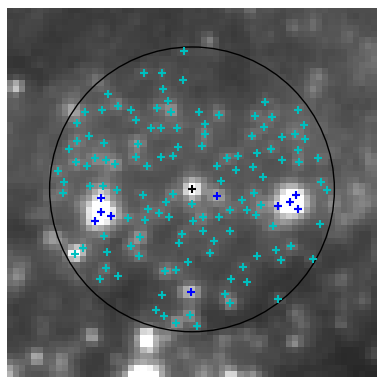}
      \caption{Spatial distribution of the stellar sources in the local environment of SN\,2023fyq, overlaid on the HST/WFC3/UVIS/F814W image. The black "+" symbol marks the position of the Source~A, and the surrounding black circle has a radius of 100~pc. The blue "+" symbols show the position of bright blue sources expected from the younger stellar population. The cyan "+" symbols denote the remaining sources that are not readily classified, probably including a mix of younger and older populations.}
         \label{fig:cmdstars}
   \end{figure}
   
To test whether the two populations are spatially clustered, we applied two criteria to extract a clean sample of likely young blue sources: F814W~$<$~23~mag and F438W - F814W~$<$~3~mag. The remaining sources are unclassified and probably include a mix of younger and older populations. Fig.~\ref{fig:cmdstars} displays their spatial distribution, revealing that the younger population sources are mostly concentrated in two regions, implying a certain degree of spatial clustering.

\subsection{A simple picture of the SN environment \label{subsec:stru}}

In Section~\ref{subsec:gal} we have already shown the distribution of stars on the optical and near-IR images as well as the molecular gas and ionized gas in the host galaxy of SN\,2023fyq. A key phenomenon is that the spiral arms traced by stars are highly asymmetric on the optical images, whereas on the near-IR images they are much more symmetric. As previously suggested, this is most likely due to different relative distributions of stars and dust along the line of sight: dust resides in front of stars on the eastern arm and causes much higher obscuration, whereas on the western arm most dust is in the background of stars and has a much smaller influence.

In the local SN environment, we find two stellar populations (Section~\ref{subsec:cmd}). The older population has lower host extinction ($A_V^{\rm{host}}$ = 0.11~mag) compared to the younger population ($A_V^{\rm{host}}$ = 0.26~mag). In contrast, the ionized gas shows a noticeably higher extinction ($A_V^{\rm{host}}$ = 0.8~mag) than both stellar components (Section~\ref{subsec:gal}; Fig.~\ref{fig:galaxy}g).
This indicates that along the line of sight at the SN site, the components ordered from near to far are the older stellar population, the younger stellar population, and ionized gas, so that they have increasingly higher interstellar extinctions. 
The molecular gas (Fig.~\ref{subsec:gal}c) is most likely to reside in the background of the ionized gas since otherwise the large amount of dust in the molecular clouds would obscure the light of the other components.

\begin{figure}[ht!]
   \centering
   \includegraphics[width=\hsize]{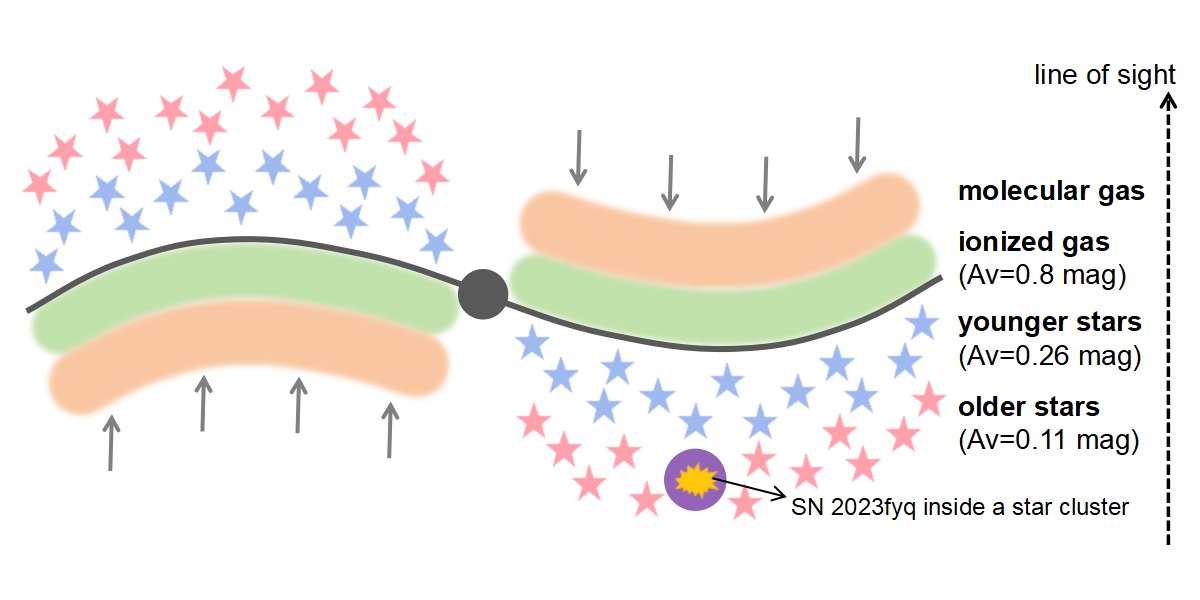}
      \caption{Schematic diagram illustrating the structure of SN\,2023fyq's host galaxy, NGC\,4388, as viewed face-on. The gray arrows show the direction of gas motion.}
         \label{fig:stru}
   \end{figure}

Based on the above discussion, we suggest a 3D structure of the host galaxy as illustrated by Fig.~\ref{fig:stru}. This 3D structure is consistent with a scenario in which spiral arms trigger sequential star formation as gas propagates through its potential well. Note that on the western arm, gas is moving toward the observer along the line of sight (Fig.~\ref{fig:galaxy}e) and that SN\,2023fyq, at a distance of only $\sim$0.9\,kpc from the galaxy center, is most likely to reside within the spiral arm-disk corotation radius (which is typically $\gtrsim$~4~kpc for a NGC4388-like galaxy; \citealp{2020Abdeen, 2024Marchuk, 2024Kostiuk, 2025Kostiuk}). Therefore, the molecular gas near the SN site moves faster than the spiral arm and it propagates through the spiral arm from behind; during this process the gas gets compressed, forming stars, and is then ionized by the ultraviolet radiation of the newly formed stars. This mechanism leads to a sequence of molecular gas, ionized gas, younger stars, and older stars at different positions relative to the spiral arms, as shown in Fig.~\ref{fig:stru} for NGC\,4388 and often seen in many other galaxies (\citealp{2015Hou}). 
Although this structure may not be entirely accurate, it appears to be reasonably consistent with the available observations.

In Section~\ref{sec:progenitor} we detected a pre-explosion source at the SN position and modeled its SED as a SN progenitor with a star cluster. 
The age of the star cluster derived from the SED fitting is log($t$/yr) = 7.1~$\pm$~0.1, which is very close to the age of the older stellar population (log($t$/yr) = $7.22^{+0.02}_{-0.02}$) obtained by fitting the resolved stellar populations in the local SN environment (Section~\ref{subsec:cmd}).
Regarding the SN progenitor, \citet{2024Dong} have obtained a host-galaxy reddening of $E(B-V)_{\rm host} = 0.037\pm0.09$\,mag, or an extinction of $A_V^{\rm host}=0.11\pm0.28$\,mag, using Na\,\textsc{i}\,\textsc{d} absorption doublets. 
This value is consistent with the extinction of the older stellar population in the local SN environment ($A_V^{\rm host}$ = 0.11~mag) and is lower than that of the younger population ($A_V^{\rm host}$ = 0.26~mag), although the large error makes this difference statistically less robust. 
Additionally, the spatial distribution of the stellar populations (see Fig.~\ref{fig:cmdstars}) reveals that the younger stellar population exhibits some degree of clustering, whereas our detected source at the SN position does not fall within those clustered regions.
Therefore, it is plausible that both the SN progenitor and the star cluster are members of the older population, so that they share similar ages, interstellar extinctions, and spatial distributions; it is also possible that the star cluster hosts the SN progenitor, although chance alignment cannot be ruled out. Based on the above discussion, the age of the progenitor of SN\,2023fyq is estimated to be log($t$/yr) = 7.1--7.2.

\section{Summary and discussion \label{sec:sum}}

SN\,2023fyq is a Type~Ibn SN in the nearby galaxy NGC\,4388. In this paper, we performed a detailed analysis of its progenitor and environment based on HST and JWST imaging, VLT/MUSE IFU spectroscopy, and ALMA CO(2--1) radio interferometry.

On the pre-explosion HST images, we detected a point source at the SN position in all the observed bands from F336W to F160W. Its brightness in F336W significantly faded from pre‑explosion to late times, confirming that this source is physically associated with the SN. We attribute this dimming to an SN progenitor that has disappeared and the remaining late-time brightness to a star cluster. On this assumption, the pre-explosion source can be modeled as the combination of a hot progenitor, with a temperature of $>$15000~K, a luminosity of log(L/L$_\odot$)~$\gtrsim$~5.5, and a radius of log(R/R$_\odot$)~$\lesssim$~2, and a star cluster with an age of log($t$/yr) = 7.1~$\pm$~0.1. This makes SN\,2023fyq the first Type~Ibn SN with a directly detected progenitor and a possible host star cluster.
Note that if any SN light remained at late epochs, it would imply that the cluster luminosity is overestimated, which in turn would lead to a higher progenitor luminosity and an older cluster age.

Meanwhile, we carried out a detailed analysis of the SN host galaxy by probing the stars, ionized gas, and molecular gas with the multiwavelength dataset. In the local SN environment, we found two stellar populations with different ages and extinctions. Based on these observations, we suggest a 3D structure of the host galaxy, in which spiral arms trigger sequential star formation as gas propagates through its potential well. The SN progenitor that we have identified in the pre-explosion images is most likely to be associated with the older stellar population and have an age of log($t$/yr) = 7.1--7.2.
This age corresponds to an initial mass of 13.7--16.2~$M_\odot$ for a single star. Given the cluster mass of log($M$/$M_\odot$)= $3.8^{+0.1}_{-0.1}$, the cluster is massive enough to host such a star.

In the HR diagram, the position of SN\,2023fyq's progenitor is consistent with a single very massive star of $>$100~$M_\odot$ (Fig.~\ref{fig:hr}). 
However, we deem a single very massive star progenitor very unlikely for SN\,2023fyq for two primary reasons:  
(1) the estimated ejecta mass is only 0.54~$M_\odot$ for SN\,2023fyq \citep{2024Dong}, significantly lower than expected for a single very massive progenitor;
(2) the derived progenitor age from its environment is relatively old and much longer than the typical evolutionary timescale for such very massive stars.

\cite{2024Dong} proposed a binary progenitor system for SN\,2023fyq consisting of a low-mass helium star ($\sim$2.5--3~$M_{\odot}$) and a compact companion. In their model, mass transfer via Roche-lobe overflow forms an equatorial disk ($\sim$0.6~$M_{\odot}$) and an extended envelope ($\sim$0.3~$M_{\odot}$). 
The precursor emission that they have detected is likely to arise from the collision between a disk wind, launched by super-Eddington accretion onto the compact companion, and external material formed from previous mass-transfer outflow. 
This helium star + compact companion model is also supported by \cite{2024Tsuna} and \cite{2025Baer-Way}.
In this scenario, the emission of our detected SN progenitor may arise from the same mechanism, i.e., from the collision between a disk wind and external material.
Our result also suggests that precursor emission from the binary interaction had already started as early as $\sim12$~yr before the SN explosion, long before the first precursor detection by ground-based telescopes.

SN\,2023fyq adds to the diversity of Type~Ibn SNe in terms of their progenitor channels and mass-loss mechanisms.
For example, the binary companion of the Type~Ibn SN\,2006jc, discovered by \cite{2016Maund} and \cite{2020Sun}, is a non-degenerate supergiant, while that of SN\,2023fyq is likely a compact object \citep{2024Dong, 2024Tsuna, 2025Baer-Way}. This suggests that their progenitor systems have very different parameters and have undergone very different evolutionary paths.
Similarly, the CSM of SN\,2006jc was formed via a giant eruption of the progenitor (clearly detected in the two years before explosion; \citealp{2007Foley, 2007Pastorello}) and that of SN\,2023fyq was formed via more steady binary mass transfer. 
These differences indicate that Type~Ibn SNe may have remarkable diversities in both progenitor channels and mass-loss mechanisms, the details of which still await further investigations.

\begin{acknowledgements}
We are grateful to Dr. C{\'e}sar Rojas-Bravo for his helpful comments on this paper.
This work is supported by the National Natural Science Foundation of China Grant No.12303051, the Strategic Priority Research Program of the Chinese Academy of Sciences Grant No. XDB0550300, and by the China Manned Space Program Grant No. CMS-CSST-2025-A14. Y.S. acknowledges support from the National Natural Science Foundation of China (NSFC) with grant No. 12303001, and the Fundamental Research Funds for the Central Universities.
This work is based on observations made with the NASA/ESA Hubble Space Telescope (Programs 12365, 12185 and 17535) and observations made with the NASA/ESA/CSA James Webb Space Telescope (Programs 2064, 6659 and 5627). The data were obtained from the Space Telescope Science Institute, which is operated by the Association of Universities for Research in Astronomy, Inc., under NASA contract NAS 5–26555 for HST and NASA contract NAS 5-03127 for JWST.
This work is also based on observations collected at the European Southern Observatory under ESO programme 108.229J.001 and data obtained from the ESO Science Archive Facility with DOI under https://doi.org/10.18727/archive/42.
Additionally, this work makes use of the following ALMA data: ADS/JAO.ALMA\#2019.1.00763.L, ADS/JAO.ALMA\#2023.1.00026.S. ALMA is a partnership of ESO (representing its member states), NSF (USA) and NINS (Japan), together with NRC (Canada), NSTC and ASIAA (Taiwan), and KASI (Republic of Korea), in cooperation with the Republic of Chile. The Joint ALMA Observatory is operated by ESO, AUI/NRAO and NAOJ.
\end{acknowledgements}

\bibliographystyle{aa}
\bibliography{cite}

\begin{appendix}

\onecolumn

\section{HST and JWST observations \label{app:obs}}
Table~\ref{tab:data} lists detailed information on the HST and JWST observations of SN\,2023fyq used in this work. 

\begin{table*}[ht!]
\centering
\caption{HST and JWST observations of SN\,2023fyq used in this work. \label{tab:data}}
    \begin{tabular}{ccccccc}
    \hline
    \hline
    
     Telescope & Proposal & Date & Epoch\tablefootmark{a} &  Instrument & Filter & Exposure  \\
     & ID & & (day) & & & Time (s) \\
    
    \hline
    HST  & 12365 & 2011-05-11 & $-$4461 & WFC3/UVIS & FQ508N & 574 \\
                && 2011-05-11 & $-$4461 & WFC3/UVIS & F665N  & 474 \\
                && 2011-05-11 & $-$4461 & WFC3/UVIS & F547M  & 194 \\
                && 2011-05-11 & $-$4461 & WFC3/UVIS & F621M  & 194 \\
                && 2011-05-11 & $-$4461 & WFC3/UVIS & F673N  & 306 \\
           
         & 12185 & 2011-06-08 & $-$4433 & WFC3/IR   & F110W & 147 \\
                && 2011-06-08 & $-$4433 & WFC3/IR   & F160W & 422 \\
                && 2011-06-08 & $-$4433 & WFC3/UVIS & F336W & 1230 \\
                && 2011-06-08 & $-$4433 & WFC3/UVIS & F438W & 410 \\
                && 2011-06-08 & $-$4433 & WFC3/UVIS & F814W & 1952 \\
    
         & 17535 & 2025-03-07 & 588 & WFC3/UVIS & F336W & 2328 \\
          
    \hline
    JWST & 2064 & 2023-06-27 & $-$31 & MIRI & F1800W & 358 \\
         & 6659 & 2024-06-13 & 321 & MIRI & F1800W & 358 \\
         & 5627 & 2025-01-10 & 532 & NIRCAM & F182M & 1632 \\
               && 2025-01-10 & 532 & NIRCAM & F335M & 1632 \\
                
    \hline\hline
 \end{tabular}

\tablefoot{PIs: 12365: Wang; 12185: Greene; 17535: Veilleux; 2064: Rosario; 6659: Taggart; 5627: Veilleux. }
\tablefoottext{a}{The epochs are with respect to the maximum light in the $r$ band at JD = 2,460,154.3 $\pm$ 0.5, as reported by \cite{2024Dong}.}

\end{table*}

\FloatBarrier 

\section{Decontamination of neighboring sources \label{app:mixed}}
\begin{figure}[ht!]
   \centering
   \includegraphics[width=0.7\hsize]{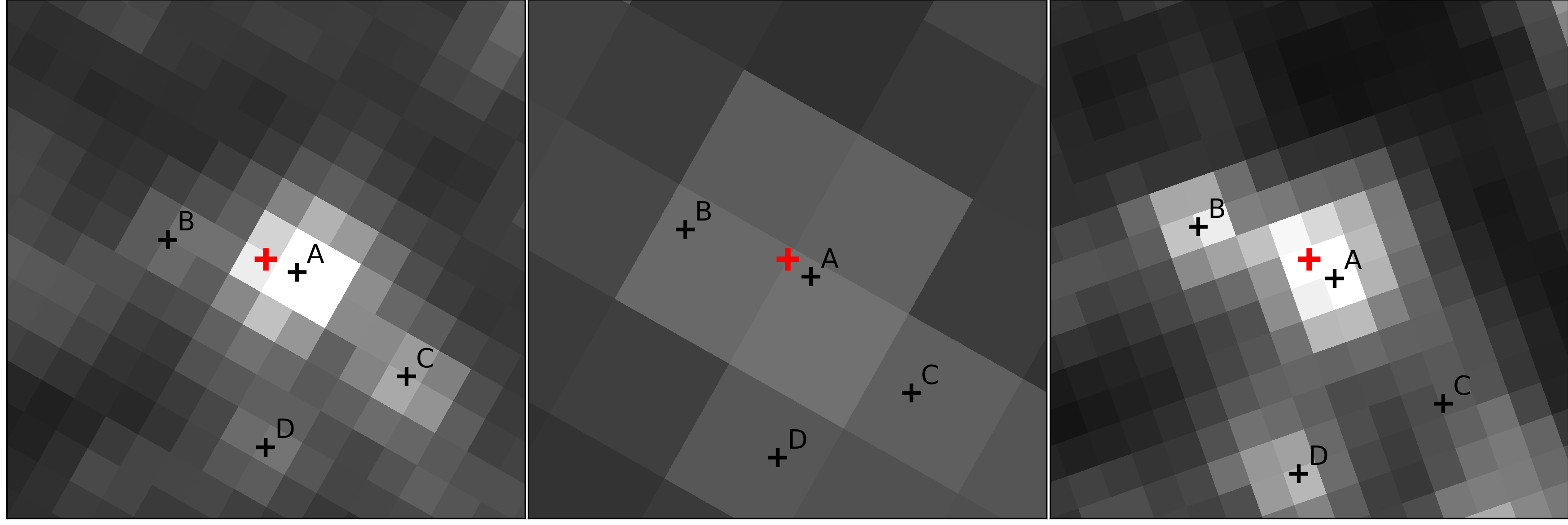}
      \caption{Images observed by HST/WFC3/UVIS/F814W (left), HST/WFC3/IR/F110W (middle) and JWST/NIRCam/F182M (right) near the SN site. The red "+" symbols are the photo-center of the SN progenitor detected on HST/WFC3/IR/F110W image. The black "+" symbols are the centers of the Sources A, B, C and D detected on the other two higher-resolution images. We used common stars to compute the coordinate transformation matrices and mark the above source positions on all panels.}
         \label{fig:mixed}
   \end{figure}

Since the HST/WFC3/IR images have a relatively lower spatial resolution, the SN progenitor may be blended with close neighboring stars and its photometry could be contaminated. As shown in Fig.~\ref{fig:mixed}, the pixel size of HST/WFC3/IR (0.12") is much larger than those of HST/WFC3/UVIS and JWST/NIRCam (0.04" and 0.031", respectively).
On the higher-resolution HST/WFC3/UVIS and JWST/NIRCam images, four distinct sources can be detected at or near the SN position (Sources A, B, C and D). Using coordinate transformation matrices, we register the positions of these sources on all observations. We find that on the lower-resolution HST/WFC3/IR images, while Source~C and Source~D can still be marginally resolved, Souce~A and Source~B are completely blended and the photocenter of the blended source lies between the two sources. Therefore, photometry of the SN progenitor is mainly affected by the contamination of Source~B. Accordingly, we perform a linear interpolation of the flux of Source~B across the F814W and F182M bands to estimate its brightness in the F110W and F160W bands and we consider two extremes, in which Source~B contributes no or all of its brightness to the total photometry of the SN progenitor. In this way, we can estimate and remove the contamination in the photometry and the range spanned by the two extremes is taken as the error bar associated with the data points in Fig.~\ref{fig:sed}).

\FloatBarrier 

\section{VLT/MUSE spectral line fitting \label{app:specline}}
\begin{figure}[ht!]
   \centering
   \includegraphics[width=0.24\hsize]{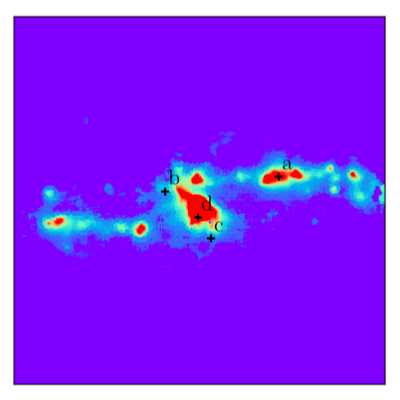}
   \includegraphics[width=0.74\hsize]{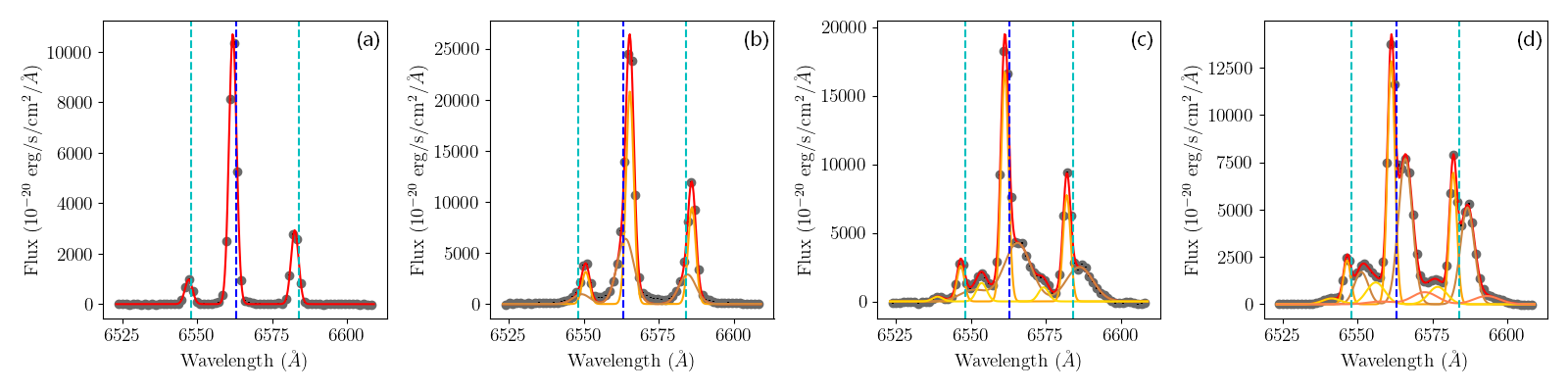}
      \caption{The wavelength-integrated flux map of the H$\alpha$ line total emission observed by VLT/MUSE and spectral line fitting results with 1, 2, 3, 4 components corresponding to the four positions marked by the "+" symbols (i.e. a, b, c and d, respectively).}
         \label{fig:Itotal}
   \end{figure}

Since NGC\,4388 is an edge-on galaxy, the substantial line-of-sight superposition of distinct gaseous components across the galaxy gives rise to complex emission-line profiles in our measurements. Emission lines along the spiral arms typically display a single kinematic component, while those in the nuclear region exhibit 2--4 distinct components. We present four example regions with different numbers of spectral line components; their positions within the host galaxy and the corresponding Gaussian fitting results for the emission lines are shown in Fig.~\ref{fig:Itotal}.

\FloatBarrier 

\section{BPT diagram \label{app:bpt}}
\begin{figure}[ht!]
   \centering
   \includegraphics[width=0.4\hsize]{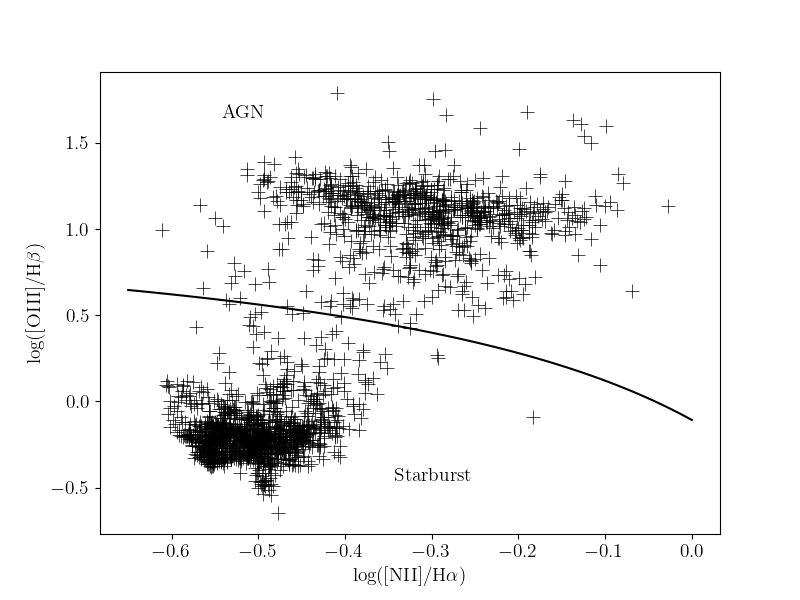}
      \caption{BPT diagram of NGC\,4388. The black line is the maximum starburst line defined by \cite{2001Kewley}.}
         \label{fig:bpt}
   \end{figure}

Fig.~\ref{fig:bpt} shows the BPT diagram \citep{1981Baldwin} with the maximum starburst line defined by \cite{2001Kewley}. 
The BPT diagram is a powerful optical diagnostic tool to distinguish the dominant ionization mechanism of ionized gas. With this diagram, we found that the spiral arms are ionized by star formation while the nuclear region is ionized by AGN.

\FloatBarrier 
\clearpage

\end{appendix}
\end{document}